\documentclass[11pt]{article}
\usepackage{times}
\usepackage[tbtags]{amsmath}
\usepackage{amsthm}
  \theoremstyle{plain}
  \newtheorem{theorem}{Theorem}
  \newtheorem{lemma}[theorem]{Lemma}

  \theoremstyle{definition}
  \newtheorem{definition}[theorem]{Definition}

  \theoremstyle{remark}

  \theoremstyle{plain}
  \newtheorem*{theorem*}{Theorem}
  \newtheorem*{lemma*}{Lemma}
  \newtheorem*{corollary*}{Corollary}
  \newtheorem*{proposition*}{Proposition}
  \newtheorem*{claim*}{Claim}
  
\usepackage{amssymb}
\usepackage{amsfonts}

\newlength{\actualtopmargin}
\newlength{\actualsidemargin}
\newcommand{\bbC}{\mathbb{C}}

\newcommand{\bbF}{\mathbb{F}}

\newcommand{\bbZ}{\mathbb{Z}}

\newcommand{\bmb}{\boldsymbol{b}}

\newcommand{\bmv}{\boldsymbol{v}}

\newcommand{\bmx}{\boldsymbol{x}}
\newcommand{\bmy}{\boldsymbol{y}}
\newcommand{\bmz}{\boldsymbol{z}}

\newcommand{\bmzero}{\boldsymbol{0}}
\newcommand{\bmone}{\boldsymbol{1}}

\newcommand{\calH}{\mathcal{H}}

\newcommand{\sfA}{\mathsf{A}}
\newcommand{\sfB}{\mathsf{B}}

\newcommand{\sfR}{\mathsf{R}}
\newcommand{\sfS}{\mathsf{S}}
\newcommand{\sfT}{\mathsf{T}}

\newcommand{\CNOT}{\mathrm{CNOT}}

\newcommand{\bra}[1]{\langle #1 \vert}

\newcommand{\ket}[1]{\vert #1 \rangle}

\newcommand{\ketbra}[1]{\vert #1 \rangle \langle #1 \vert}

\newcommand{\abs}[1]{\vert #1 \vert}

\newcommand{\size}[1]{\vert #1 \vert}

\newcommand{\function}[3]{{#1 \colon #2 \rightarrow #3}}

\newcommand{\Complex}{\bbC}

\newcommand{\Integers}{\bbZ}

\newcommand{\Field}{\bbF}
\newcommand{\Binary}{{\{ 0, 1 \}}}

\begin{document}

\sloppy


\title{\Large
  \textbf{
    Perfect Quantum Network Communication Protocol\\
    Based on Classical Network Coding
  }\\
}

\author{
  Hirotada Kobayashi\footnotemark[1]~~\footnotemark[2]\\
  \and
  Fran\c{c}ois Le Gall\footnotemark[2]\\
  \and
  Harumichi Nishimura\footnotemark[3]\\
  \and
  Martin R\"otteler\footnotemark[4]\\
}

\date{}

\maketitle
\thispagestyle{empty}
\pagestyle{plain}
\setcounter{page}{0}

\vspace{-5mm}

\renewcommand{\thefootnote}{\fnsymbol{footnote}}

\begin{center}
{\large
  \footnotemark[1]%
  Principles of Informatics Research Division\\
  National Institute of Informatics, Tokyo, Japan\\
  [2.5mm]
  \footnotemark[2]%
  Quantum Computation and Information Project\\
  Solution Oriented Research for Science and Technology\\
  Japan Science and Technology Agency, Tokyo, Japan\\
  [2.5mm]
  \footnotemark[3]%
  Department of Mathematics and Information Sciences\\
  Graduate School of Science\\
  Osaka Prefecture University, Sakai, Osaka, Japan\\
  [2.5mm]
  \footnotemark[4]%
  NEC Laboratories America, Inc., Princeton, NJ, USA
}\\
[5mm]
{\large 8 February 2009}\\
[8mm]
\end{center}



\begin{abstract}
  This paper considers a problem of quantum communication between parties that
  are connected through a network of quantum channels.
  The model in this paper assumes that there is no prior entanglement
  shared among any of the parties, but that classical communication is free.
  The task is to perfectly transfer an unknown quantum state
  from a source subsystem to a target subsystem,
  where both source and target are formed by ordered sets of some of the nodes.
  It is proved that a lower bound of the rate
  at which this quantum communication task is possible
  is given by the classical min-cut max-flow theorem of network coding,
  where the capacities in question
  are the quantum capacities of the edges of the network. 
\end{abstract}

\clearpage


\section{Introduction}

Consider a communication network consisting of a set~$V$ of several nodes,
each of which can hold a small number of qubits
and which have no prior entanglement among them.
Furthermore, these nodes are connected via a set~$E$ of edges
which correspond to quantum communication channels, each of a certain capacity.
Let ${G = (V,E)}$ be the weighted graph corresponding to this network.
Consider the following communication problem:
given a set~${S \subseteq V}$ of source nodes in the network
which all together hold a quantum state~$\rho_S$
and a set of target nodes~${T \subseteq V}$ of nodes
to which the quantum state is supposed to be sent,
where ${\size{S} \leq \size{T}}$,
the task is to devise a communication protocol that,
for any selected subset~${T_0 \subseteq T}$ with ${\size{T_0} = \size{S}}$,
transmits the state~$\rho_S$ through the network
such that after the transmission the state of the system
corresponding to $T_0$ is equal to $\rho_S$
and for any particular ordering of the elements of $T_0$.

Clearly, this task depends on the particular properties of the network
and it might or might not be possible to achieve this task
for the given $G$, $S$, and $T$.
A trivial case where it is impossible to transmit any state perfectly is
when $S$ and $T$ are disconnected,
i.\,e., there is no quantum communication path
between any node of $S$ and any node of $T$.
Another trivial case where it is possible to transmit any state perfectly
is when each node in $S$ is directly connected with each node in $T$.
We shall be concerned with cases in between these two extremes,
where the actual network topology given by $G$
does not allow disjoint paths between the qubits in $S$ and the qubits in $T$,
but we nevertheless want to achieve perfect state transfer
via quantum teleportation~\cite{BBC+:1993}.
If perfect state transfer is possible,
we also want to achieve it with as few uses of the network as possible.

If $G$ is a classical network,
a celebrated result of network coding
is the min-cut max-flow theorem for network information flow~\cite{ACLY:2000,KM:2003,LYC:2003}
which states that perfect transfer from $S$ to $T$ at rate~$h$ is possible
whenever for each~${t \in T}$ the max-flow between $\sigma$ and $t$
is at least~$h$.
Here $\sigma$ is a special source node ${\sigma \notin S}$
from which the input information is supposed to originate and is passed to $S$.
This is the so-called multi-cast model
for which optimal network coding is linear~\cite{LYC:2003}
and can be constructed in polynomial time~\cite{JSC+:2005}. 
This is in contrast to the general network model
in which linear coding is not enough~\cite{DFZ:2005}.  


The strategy this paper presents to achieve perfect quantum teleportation
through the quantum network $G$ is very simple
and works whenever the associated classical multi-casting task is feasible.
It consists of five steps:
(i) First, a state~${\ket{0}+\ket{1}}$ (normalization omitted)
is created at each node~${s_i \in S}$, ${1 \leq i \leq \size{S}}$.
(ii) Next, a classical linear network coding protocol for $G$, $S$, $T$
is translated into a sequence of Clifford operations to be applied
at each node of the network.
It is proved that
the states can be sent through the network in such a way
that the final state is given by $\size{S}$ cat states each of the form
${
  \ket{0}_{\sfS_i} \ket{0}_{\sfT_{1,i}} \cdots \ket{0}_{\sfT_{\size{T},i}}
  +
  \ket{1}_{\sfS_i} \ket{1}_{\sfT_{1,i}} \cdots \ket{1}_{\sfT_{\size{T},i}}
}$,
albeit some of the phases in this state might be incorrect.
Here, for each ${1 \leq i \leq \size{S}}$,
$\sfS_i$ is the single-qubit register possessed by the node~$s_i$
and each $\sfT_{j,i}$ is the single-qubit register possessed
by the node~${t_j \in T}$, ${1 \leq j \leq \size{T}}$.
(iii) Now the classical information obtained by measuring internal network qubits in the Hadamard basis
is sent to one dedicated output node~${t_1 \in T}$.
Using this information, the phase errors are fixed
and indeed $\size{S}$ perfect cat states are generated.
(iv) After the selection of ${T_0 \subseteq T}$ is revealed,
the cat states are converted into $\size{S}$ EPR pairs
shared between the corresponding node pairs.
For this purpose, it is again necessary to measure in the Hadamard basis and exchange the obtained classical information.
(v) Finally, using the EPR pairs the state~$\rho_S$ over $S$ is teleported
to the target nodes in $T_0$.

It is perhaps interesting to note that deciding what the target nodes~$T_0$ are
(and in particular their order!) to which the state is teleported
can be done \emph{after} the quantum network has been used.
At this point the only required communication is purely classical.

\paragraph{Related work} 
It should be noted that, prior to this work,
several papers studied the problem of sending quantum states
using the idea of network coding, that is, allowing any coding
at intermediate nodes of the network.
Hayashi,~Iwama,~Nishimura,~Raymond,~and~Yamashita~\cite{HIN+:2007} showed
that network coding (without free classical communication)
does not give us any benefit for perfect transmission
on the butterfly network, a famous network with two source-target pairs.
Leung,~Oppenheim,~and~Winter~\cite{LOW:2006} showed that
this negative result can be generalized to several types of networks
even if the transmission is allowed to be asymptotically perfect.
Also, they studied several variants of situations
including the one where free classical communication is allowed.
On the contrary, Hayashi~\cite{H:2007} showed that perfect transmission
of two source states on the butterfly network
can be efficiently done by network coding
if the sources have prior entanglement and each link
has a capacity of one qubit or two classical bits.
It should be noted that all of the above results
focus on the (multiple-source) uni-cast model,
a well-studied network coding model, 
while the model discussed in this paper is close to the multi-cast model.
The quantum network coding for the multi-cast model was previously studied
by Shi~and~Soljanin~\cite{SS:2006}.
In their model, however, the source was restricted to
the product of copies of a state,
and hence in fact they could use only source coding for perfect transmission,
instead of coding at intermediate nodes.


\section{Preliminaries}

\subsection{Quantum Information}
Quantum states are normalized vectors in a complex Hilbert space~${\calH = \Complex^d}$.
The simplest case of ${\calH = \Complex^2}$ is of particular importance,
and a system supporting such a state space is called a \emph{qubit (quantum bit)}.
This paper mainly treats the case of two-dimensional quantum systems,
but the results in this paper can be generalized to any $d$-dimensional systems.
Notice that even if the quantum information to be transmitted is originally
given by qubits,
higher-dimensional systems may be necessary in the coding schemes. 
Intuitively, this is because the protocols to be presented
are based on classical network coding
which itself might require higher alphabets for the coding, 
even if the original information is binary.
These points will be discussed further in Theorem~\ref{theorem2}.

The orthonormal basis states of a qubit are written as $\ket{0}$ and $\ket{1}$,
and the general state of a qubit is given by
${\ket{\phi} = \alpha\ket{0} + \beta\ket{1}}$,
where ${\alpha, \beta \in \Complex}$ and ${\abs{\alpha}^2 + \abs{\beta}^2 = 1}$.
If both $\alpha$ and $\beta$ are non-zero,
the state $\ket{\phi}$ is a so-called superposition of $\ket{0}$ and $\ket{1}$
with amplitudes $\alpha$ and $\beta$.
For a $d$-dimensional system,
we label the orthonormal basis states
by the elements of some alphabet of size $d$,
e.\,g., the numbers $\{0,1,\ldots,d-1\}$ or the elements of a finite field,
if $d$ is a prime power.
A normalized vector in $\Complex^d$ is called a \emph{qudit},
and is written as ${\ket{\psi} = \sum_{i=0}^{d-1} \alpha_i \ket{i}}$,
where ${\alpha_i \in \Complex}$ and ${\sum_{i=0}^{d-1} \abs{\alpha_i}^2 = 1}$.
\emph{Quantum registers} consist of several qudits.
The basis states of a quantum register of $n$ qudits
are tensor products of the basis states of the single qudits.
The following notation is used:
\[
\ket{x_1} \otimes \ket{x_2} \otimes \cdots \otimes \ket{x_n}
=
\ket{x_1} \ket{x_2} \cdots \ket{x_n}
=
\ket{x_1, x_2, \ldots, x_n},
\]
where ${x_1, \ldots, x_n}$ are elements of $\{0,1,\ldots,d-1\}$.
From now we focus on the case where ${d=2}$.
A general state of a quantum register of $n$ qubits
is a normalized vector in
${\calH = (\Complex^2)^{\otimes n} \cong \Complex^{2^n}}$,
given by
${\ket{\psi} = \sum_{\bmx \in \Field_2^n} \alpha_{\bmx} \ket{\bmx}}$,
where ${\alpha_{\bmx} \in \Complex}$
and ${\sum_{\bmx \in \Field_2^n} \abs{\alpha_{\bmx}}^2 = 1}$.
For two vectors $\bmx$ and $\bmy$ in $\Field_2^n$,
let ${\bmx \cdot \bmy}$ denote the usual inner product. 
When writing states of quantum registers, normalization factors may be omitted.
We next discuss some basic elementary quantum operations
that can be used to manipulate the content of quantum registers.
This is all standard, see for example~\cite{NC:2000}.

\begin{definition}[Elementary Clifford Operations]
\label{def:elemGates}
The following four operations are called elementary Clifford operations:
\begin{align*}
\sigma_X
&
:= \sum_{x \in \Field_2} \ket{x+1}\bra{x},
\\
\sigma_Z
&
:= \sum_{x \in \Field_2} (-1)^x \ketbra{x},
\\
H
&
:=
\frac{1}{\sqrt{2}} \sum_{x,y \in \Field_2} (-1)^{xy} \ket{y}\bra{x},
\\
\CNOT^{(\sfA,\sfB)}
&
:= \sum_{x,y \in \Field_2}\ketbra{x}_{\sfA} \otimes \ket{x+y}\bra{y}_{\sfB}.
\end{align*}
\end{definition}
Here, when writing $(-1)^x$ for ${x \in \Field_2}$,
we identify $\Field_2$ and ${\Integers/2\Integers}$, the integers modulo $2$. 
The operation~$\sigma_X$ corresponds to the addition of the identity element.
The operation~$\sigma_Z$ has no direct classical analogue
and changes the phases of the basis states.
The operator~$H$ is called the \emph{Hadamard operator},
and $\CNOT$ the \emph{controlled-NOT operator}. 

Finally, let
\begin{align*}
\ket{+}
&
= H \ket{0} = \frac{1}{\sqrt{2}}(\ket{0}+\ket{1}),
\\
\ket{-}
&
= H \ket{1} = \frac{1}{\sqrt{2}}(\ket{0}-\ket{1}),
\end{align*}
and for convenience,
we say that measuring a qubit in the basis $\{\ket{+},\ket{-}\}$
(the \emph{Hadamard basis}) gives a bit $b$,
where ${b = 0}$ if $\ket{+}$ is measured, and ${b = 1}$ if $\ket{-}$ is measured.

\subsection{Convention on Classical Multi-Cast}
\label{subsection:multicast}
The key result of this paper is
a quantum simulation of any classical linear network coding scheme
in the multi-cast model.
Here we use the standard definition of classical linear network coding (see~\cite{LYC:2003,JSC+:2005}).
For convenience, the following simple but very useful convention is assumed when describing a classical multi-cast (linear) protocol. 

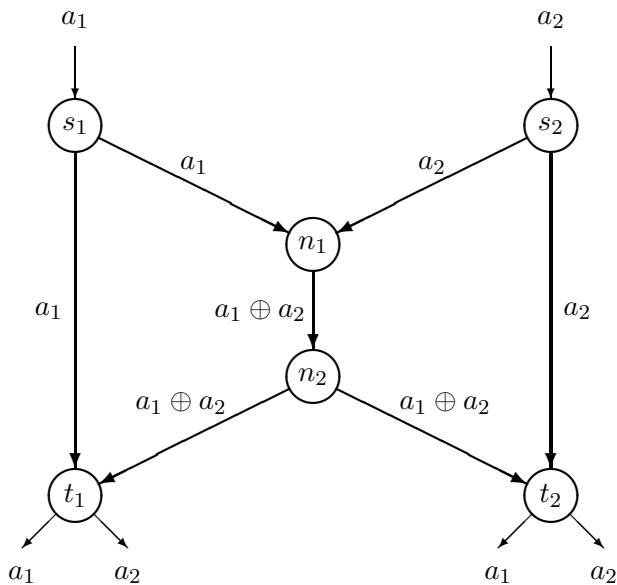
\begin{figure}[t]
\begin{center}
\begin{picture}(220,220)
\thinlines
\put( 20, 220){\makebox(0,0){$a_1$}}
\put(200, 220){\makebox(0,0){$a_2$}}
\put( 20, 210){\vector(0,-1){20}}
\put(200, 210){\vector(0,-1){20}}
\thicklines
\put( 20, 180){\circle{20}}
\put( 20, 180){\makebox(0,0){$s_1$}}
\put( 20, 170){\vector(0,-1){120}}
\put( 10, 110){\makebox(0,0){$a_1$}}
\put( 28.94, 175.53){\vector(2,-1){72.12}}
\put( 65, 165){\makebox(0,0){$a_1$}}
\put(200, 180){\circle{20}}
\put(200, 180){\makebox(0,0){$s_2$}}
\put(200, 170){\vector(0,-1){120}}
\put(210, 110){\makebox(0,0){$a_2$}}
\put(191.06, 175.53){\vector(-2,-1){72.12}}
\put(155, 165){\makebox(0,0){$a_2$}}
\put(110, 135){\circle{20}}
\put(110, 135){\makebox(0,0){$n_1$}}
\put(110, 125){\vector(0,-1){30}}
\put( 90, 110){\makebox(0,0){${a_1 \oplus a_2}$}}
\put(110,  85){\circle{20}}
\put(110,  85){\makebox(0,0){$n_2$}}
\put(101.06,  80.53){\vector(-2,-1){72.12}}
\put( 60,  75){\makebox(0,0){${a_1 \oplus a_2}$}}
\put(118.94,  80.53){\vector(2,-1){72.12}}
\put(160,  75){\makebox(0,0){${a_1 \oplus a_2}$}}
\put( 20,  40){\circle{20}}
\put( 20,  40){\makebox(0,0){$t_1$}}
\put(200,  40){\circle{20}}
\put(200,  40){\makebox(0,0){$t_2$}}
\thinlines
\put( 12.99,  32.99){\vector(-1,-1){12.99}}
\put(  0,  10){\makebox(0,0){$a_1$}}
\put( 27.01,  32.99){\vector( 1,-1){12.99}}
\put( 40,  10){\makebox(0,0){$a_2$}}
\put(192.99,  32.99){\vector(-1,-1){12.99}}
\put(180,  10){\makebox(0,0){$a_1$}}
\put(207.01,  32.99){\vector( 1,-1){12.99}}
\put(220,  10){\makebox(0,0){$a_2$}}
\end{picture}
\caption{
  The butterfly network and a classical linear coding protocol.
  The node $s_1$ (resp.~$s_2$) has for input a bit $a_1$ (resp.~$a_2$).
  The task is to send $a_1$ and $a_2$ to both $t_1$ and $t_2$. 
  The capacity of each edge is assumed to be one bit. 
  Our convention here is that $s_1$ (resp.~$s_2$) receives $a_1$ (resp.~$a_2$)
  through a virtual incoming edge,
  and that $t_1$ (resp.~$t_2$) has two virtual outgoing edges
  through which it should output $a_1$ and $a_2$, respectively.
  \label{fig:classicalbutterfly}
}
\end{center}
\end{figure}

Each source ${s_i \in S}$ is supposed to have a ``virtual'' incoming edge
from which it receives its input $a_i$.
Also, each target ${t_j \in T}$ is supposed to have $\size{S}$ ``virtual'' outgoing edges,
where $a_i$ must be output through the $i$th virtual outgoing edge,
for ${1 \leq i \leq \size{S}}$.
In this way, the source and target nodes perform 
a linear-coding operation on their inputs,
and this convention enables us to ignore the distinction between
source/target nodes and internal nodes.
These conventions are illustrated in Figure~\ref{fig:classicalbutterfly}
on the well-known coding protocol over the butterfly network.

\section{Sending Quantum States through Networks}
First, the following lemma is proved
to describe the effect of measuring in the Hadamard basis.

\begin{lemma}\label{lem:hadamardMeasurement}
  Consider a system of $n$ qubits
  and a partition of $\{1, \ldots, n\}$ into two disjoint subsets~$A$~and~$B$.
  Let $\ket{\psi_{\sfA,\sfB}}$ be a joint state given by
\[
\ket{\psi_{(\sfA,\sfB)}}
=
\sum_{\bmx \in \Field_2^n} \alpha_{\bmx} \ket{f(\bmx)}_{\sfA} \ket{g(\bmx)}_{\sfB},
\]
where
${\alpha_{\bmx} \in \Complex}$,
$\function{f}{\Field_2^n}{\Field_2^{\size{A}}}$,
$\function{g}{\Field_2^n}{\Field_2^{\size{B}}}$,
and registers~$\sfA$~and~$\sfB$ correspond to the qubits
belonging to $A$ and $B$, respectively.
Then the state in $\sfA$ obtained from $\ket{\psi_{(\sfA,\sfB)}}$
by measuring each qubit in $\sfB$ in the~$\{\ket{+}, \ket{-}\}$ basis
has the form
\[
\ket{\psi_{\sfA}}
=
\sum_{\bmx \in \Field_2^n} (-1)^{\bmy_0 \cdot g(\bmx)} \alpha_{\bmx} \ket{f(\bmx)},
\]
where ${\bmy_0 \in \Field_2^{\size{B}}}$
is a (in general random) vector of measurement results.
\end{lemma}

\begin{proof}
Applying the Hadamard transform~%
${
  H^{\otimes \size{B}}
  =
  \frac{1}{\sqrt{2^{\size{B}}}}
  \sum_{\bmy, \bmz \in \Field_2^{\size{B}}}
    (-1)^{\bmy \cdot \bmz} \ket{\bmy} \bra{\bmz}
}$
to the qubits in $\sfB$ gives the new state 
\[
\bigl(I_{\sfA} \otimes H^{\otimes \size{B}}\bigr) \ket{\psi_{(\sfA,\sfB)}}
=
\frac{1}{\sqrt{2^{\size{B}}}}
\sum_{\bmx \in \Field_2^n}
\alpha_{\bmx} \ket{f(\bmx)}
\sum_{\bmy \in \Field_2^{\size{B}}} (-1)^{\bmy \cdot g(\bmx)} \ket{\bmy}. 
\] 
Measuring the qubits in $\sfB$ in the computational basis~$\{\ket{0},\ket{1}\}$
gives a certain result~${\bmy_0 \in \Field_2^{\size{B}}}$ 
and the state collapses to the state claimed in the lemma.
\end{proof}

The next lemma shows the way of fixing phase errors
that have happened to a state,
provided that the phase errors are of a benign type. 

\begin{lemma}\label{lem:phaseCorrections}
Let $\ket{\psi}$ be a state of the form
\[
\ket{\psi} = \sum_{\bmx \in \Field_2^n} (-1)^{L(\bmx)} \alpha_{\bmx} \ket{\bmx},
\]
where $L$ is a known linear function.
Then by applying local $\sigma_Z$ operations,
$\ket{\psi}$ can be mapped to
${\sum_{\bmx \in \Field_2^n} \alpha_{\bmx} \ket{\bmx}}$.
\end{lemma}

\begin{proof}
Note that if $\function{L}{\Field_2^n}{\Field_2}$ is linear,
then $L$ maps ${\bmx =(x_1, \ldots, x_n) \in \Field_2^n}$ to
${L(\bmx) = \bmb \cdot \bmx}$ for some fixed vector
${\bmb =(b_1, \ldots, b_n) \in \Field_2^n}$.
Further note that since $L$ is known,
the vector $\bmb$ is also known,
and therefore the operation ${\bigotimes_{i=1}^n \sigma_Z^{b_i}}$
can be applied to the state, which has the effect of canceling out the phases.
\end{proof}

Next, we present three types of operations
necessary for the network communication protocol:
\emph{quantum coding operations},
\emph{quantum fan-out operations},
and \emph{measurements}.
Quantum fan-out operations can be formally viewed as quantum coding operations,
but we deal with them separately
since no coding is actually performed.
All the operations required for the protocol are elementary Clifford operations
and a supply of ancilla states that are initialized to $\ket{0}$.

\begin{description}
\item[Quantum coding operations]
  Classical network coding protocols in general perform coding at intermediate
  nodes.
  For simplicity, consider the case where each edge has capacity one.
  It is straightforward to generalize this to the case 
  where the capacities are positive integers.
  Consider a node ${v \in V}$ with $m$-fan-in and $n$-fan-out
  performing classical linear coding.
  The node $v$ has then $m$ incoming edges,
  each one conveying an element of $\Field_2$
  and labeled with a vector ${\bmv_i \in \Field_2^{\size{S}}}$,
  for ${i = 1, \ldots ,m}$.
  The outputs of the node are $n$ elements ${w_j \in \Field_2}$
  for ${j = 1, \ldots, n}$
  that are computed as suitable linear combinations
  ${w_j = \sum_{i=1}^m \gamma_{i,j} \sum_{k=1}^{\size{S}} v_{i,k}}$,
  where $v_{i,k}$ denotes the $k$th entry of $\bmv_i$,
  and are further propagated through the network.
  Here $\gamma_{i,j}$ are fixed elements of $\Field_2$.
  The quantum coding operation associated with this classical operation is
  as follows:
  attach $n$ new ancilla qubits initialized to $\ket{0}$
  and, for each ${i = 1, \ldots, m}$ and ${j = 1, \ldots, n}$,
  apply a controlled-NOT operation if and only if ${\gamma_{i,j} = 1}$,
  using the $i$th incoming qubit as control and the $j$th ancilla as target.
  The effect of this is to map,
  for any ${\bmx = (x_1, \ldots, x_m) \in \Field_2^m}$,
  the basis state ${\ket{\bmx} \otimes \ket{0}^{\otimes n}}$ to
  $\ket{\bmx, z_1, \ldots, z_n}$ where ${z_j = \sum_{i=1}^m \gamma_{i,j} x_i}$.
  Next, the $n$ ancilla qubits are sent along on the $n$ outgoing edges
  and all the incoming qubits are retained at the node.
\item[Fan-out operations]
  The $n$-fan-out operation is the special case of
  the quantum coding operations with one-fan-in and $n$-fan-out,
  such that ${\gamma_{1,j} = 1}$ for each ${j = 1, \ldots, n}$.
  For a given basis vector $\ket{x}$ on one qubit (with ${x \in \Field_2}$),
  we attach $n$ further ancillas initialized to $\ket{0}$
  and apply a sequence of $n$ controlled-NOT operations
  using the given qubit as control and each ancilla as target.
  The effect on the state is given by
  ${\ket{x} \ket{0}^{\otimes n} \mapsto \ket{x}^{\otimes (n+1)}}$. 
\item[Measurements]
  They are used to make the superfluous qubits (kept at each node) collapse,
  by measuring them in the Hadamard basis.
  More details will be given below in the proof of Theorem~\ref{th:quantumNC}.
\end{description}

Putting it all together, we have the following result:
\begin{theorem}
\label{th:quantumNC}
  Let ${G = (V,E)}$ be a quantum network with
  a subset~${S \subseteq V}$ of source nodes
  and a subset~${T \subseteq V}$ of target nodes,
  where each edge~${e \in E}$ has an integral weight
  that describes its quantum capacity.
  Assume that classical linear network coding over $\Field_2$
  is possible in the multi-cast model from $S$ to $T$.
  Then perfect quantum teleportation
  from $S$ to any ordered subset~${T_0 \subseteq T}$
  with ${\size{T_0} = \size{S}}$ is possible.
\end{theorem}

\begin{proof}
First, each node~${s \in S}$ creates
the state~${\ket{+} = \frac{1}{\sqrt{2}}(\ket{0}+\ket{1})}$.
Next, we simulate a classical coding scheme for the associated multi-cast task
in such a way that the fan-out operation is applied
whenever a broadcast is performed in the associated classical protocol
and the quantum coding operation is applied
whenever a classical coding operation is applied in the associated classical protocol.
Remember that, from the convention of Subsection~\ref{subsection:multicast},
the sources and target nodes are not necessary to be treated as special nodes. 

Because of the classical network coding property
that each output can perfectly recover all the inputs ${a_1, \ldots, a_{\size{S}}}$,
we obtain the following state
after the sequence of quantum coding and fan-out operations above:
\[
\frac{1}{\sqrt{2^{\size{S}}}}
\sum_{a_1, \ldots, a_{\size{S}} \in \Field_2}
  \underbrace{\ket{a_1, \ldots, a_{\size{S}}}}_{S}
  \otimes
  \underbrace{\ket{a_1, \ldots, a_{\size{S}}}}_{t_1}
  \otimes
  \cdots
  \otimes
  \underbrace{\ket{a_1, \ldots, a_{\size{S}}}}_{t_{\size{T}}} 
  \otimes
  \ket{f_1(a_1, \ldots, a_{\size{S}})}
  \otimes
  \cdots
  \otimes
  \ket{f_m(a_1, \ldots, a_{\size{S}})}
\]
for some functions~$\function{f_i}{\Field_2^{\size{S}}}{\Field_2}$,
${1 \leq i \leq m}$,
where the first $\size{S}$~qubits are owned by the source nodes in $S$,
the next ${\size{T} \cdot \size{S}}$~qubits are owned by the nodes ${t_1, \ldots, t_{\size{T}}}$ in $T$,
and the last $m$~qubits are owned by several nodes in the network. 
Note that by induction all functions $f_i$ are linear.
By Lemma~\ref{lem:hadamardMeasurement},
the first ${(\size{T}+1) \cdot \size{S}}$~qubits
must form the following state
after measuring all the last $m$~qubits in the Hadamard basis:
\[
\frac{1}{\sqrt{2^{\size{S}}}}
\sum_{a_1, \ldots, a_{\size{S}} \in \Field_2} 
  (-1)^{L(a_1, \ldots, a_{\size{S}})}
  \underbrace{\ket{a_1, \ldots, a_{\size{S}}}}_{S}
  \otimes
  \underbrace{\ket{a_1, \ldots, a_{\size{S}}}}_{t_1}
  \otimes
  \cdots
  \otimes
  \underbrace{\ket{a_1, \ldots, a_{\size{S}}}}_{t_{\size{T}}},
\]
where $\function{L}{\Field_2^{\size{S}}}{\Integers/2\Integers}$
is a linear function determined by the measurement results. 
Now, the information about $L$ is propagated
through (free) classical communication to one of the target nodes,
without loss of generality the first target node.
Using Lemma~\ref{lem:phaseCorrections},
node~$t_1$ can apply a local unitary operation
that fixes the phase and leads to the state
\[
\frac{1}{\sqrt{2^{\size{S}}}}
\sum_{a_1, \ldots, a_{\size{S}} \in \Field_2} 
  \underbrace{\ket{a_1, \ldots, a_{\size{S}}}}_{S}
  \otimes
  \underbrace{\ket{a_1, \ldots, a_{\size{S}}}}_{t_1}
  \otimes
  \cdots
  \otimes
  \underbrace{\ket{a_1, \ldots, a_{\size{S}}}}_{t_{\size{T}}}.
\]
This state is a collection of $\size{S}$~cat states,
each of ${\size{T}+1}$ qubits,
which are shared in such a way that each source node has one qubit 
and each target node has one qubit. 

When a subset~${T_0 \subseteq T}$ with ${\size{T_0} = \size{S}}$
and a permutation~$\pi$ over the $\size{S}$~elements of $T_0$ are revealed,
the $\size{T}$~parties run a protocol to prepare $\size{S}$~EPR pairs
from the $\size{S}$~cat states.
For this, again Lemmas~\ref{lem:hadamardMeasurement}~and~\ref{lem:phaseCorrections}
can be used to achieve the preparations of the EPR pairs
using local measurements and classical communication only.
Finally, the state~$\rho_S$ is teleported~\cite{BBC+:1993}
to the qubits in $T_0$ with the particular ordering given by $\pi$.
\end{proof}

In fact, Theorem~\ref{th:quantumNC} can be generalized to the following statement.
\begin{theorem}
\label{theorem2}
  Let ${G = (V,E)}$ be a quantum network with
  a subset~${S \subseteq V}$ of source nodes
  and a subset~${T \subseteq V}$ of target nodes,
  where each edge~${e \in E}$ has an integral weight
  that describes its quantum capacity.
  Assume that classical network coding
  is possible in the multi-cast model from $S$ to $T$.
  Then perfect quantum teleportation
  from $S$ to any ordered subset ${T_0 \subseteq T}$
  with $\size{T_0} = \size{S}$ is possible.
\end{theorem}

\begin{proof}[Proof (sketch)]
It is known~\cite{LYC:2003,JSC+:2005} that,
if classical multi-cast is feasible on a network,
a linear coding scheme exists over some large enough finite field.
The techniques developed in this section generalize to any finite field as follows.
Suppose that the finite field has size~$q$.
Each source node starts with
the $q$-dimensional quantum state
${\frac{1}{\sqrt{q}} \sum_{x \in \Field_q} \ket{x}}$. 
The node-by-node simulation of Theorem~\ref{th:quantumNC}
is then performed in a similar way.
To deal with the measurements,
we need a simple generalization of Lemmas~\ref{lem:hadamardMeasurement}~and~\ref{lem:phaseCorrections}
to $q$-dimensional quantum systems.
This can be done using the concept of $q$-ary Clifford operations
(see Refs.~\cite{Gottesman:99,GRB:2003} for a description of these operations in the framework of quantum error-correcting codes).
\end{proof}



\section{Example: The Butterfly Graph}

\begin{figure}[t]
\begin{center}
\begin{picture}(220,220)
\thinlines
\put( 20, 210){\vector( 0,-1){20}}
\put( 20, 220){\makebox(0,0){$\sfS_1$}}
\put(  0, 180){\vector( 1, 0){10}}
\put(-10, 180){\makebox(0,0){$\sfS'_1$}}
\put(200, 210){\vector( 0,-1){20}}
\put(200, 220){\makebox(0,0){$\sfS_2$}}
\put(220, 180){\vector(-1, 0){10}}
\put(230, 180){\makebox(0,0){$\sfS'_2$}}
\thicklines
\put( 20, 180){\circle{20}}
\put( 20, 180){\makebox(0,0){$s_1$}}
\put( 20, 170){\vector(0,-1){120}}
\put( 10, 110){\makebox(0,0){$\sfR_1$}}
\put( 28.94, 175.53){\vector(2,-1){72.12}}
\put( 65, 165){\makebox(0,0){$\sfR_2$}}
\put(200, 180){\circle{20}}
\put(200, 180){\makebox(0,0){$s_2$}}
\put(200, 170){\vector(0,-1){120}}
\put(210, 110){\makebox(0,0){$\sfR_3$}}
\put(191.06, 175.53){\vector(-2,-1){72.12}}
\put(155, 165){\makebox(0,0){$\sfR_4$}}
\put(110, 135){\circle{20}}
\put(110, 135){\makebox(0,0){$n_1$}}
\put(110, 125){\vector(0,-1){30}}
\put(100, 110){\makebox(0,0){$\sfR_5$}}
\put(110,  85){\circle{20}}
\put(110,  85){\makebox(0,0){$n_2$}}
\put(101.06,  80.53){\vector(-2,-1){72.12}}
\put( 65,  70){\makebox(0,0){$\sfR_6$}}
\put(118.94,  80.53){\vector(2,-1){72.12}}
\put(155,  70){\makebox(0,0){$\sfR_7$}}
\put( 20,  40){\circle{20}}
\put( 20,  40){\makebox(0,0){$t_1$}}
\put(200,  40){\circle{20}}
\put(200,  40){\makebox(0,0){$t_2$}}
\thinlines
\put( 12.99,  32.99){\vector(-1,-1){12.99}}
\put(  0,  10){\makebox(0,0){$\sfT_1$}}
\put( 27.01,  32.99){\vector( 1,-1){12.99}}
\put( 40,  10){\makebox(0,0){$\sfT'_1$}}
\put(192.99,  32.99){\vector(-1,-1){12.99}}
\put(180,  10){\makebox(0,0){$\sfT_2$}}
\put(207.01,  32.99){\vector( 1,-1){12.99}}
\put(220,  10){\makebox(0,0){$\sfT'_2$}}
\end{picture}
\caption{
  Example for perfect quantum state transfer through a quantum network.
  This example is based on the well-known butterfly network.
  Each edge has quantum capacity one.
  The task is to send a given input quantum state~$\rho_S$
  in ${(\sfS_1, \sfS_2)}$
  to either ${(\sfT_1, \sfT_2)}$ or ${(\sfT_2, \sfT_1)}$
  in this order of registers.
  Here, the quantum register~$\sfS_1$~(resp.~$\sfS_2$)
  is possessed by the source node~$s_1$~(resp.~$s_2$),
  while the quantum register~$\sfT_1$~(resp.~$\sfT_2$)
  is possessed by the target node~$t_1$~(resp.~$t_2$).
  The protocol given in Theorem~\ref{th:quantumNC} realizes
  perfect quantum teleportation of $\rho_S$
  for both possible orders of the target registers.
  Each $\sfR_i$ indicates the quantum register to be sent
  along the corresponding edge in the protocol.
  The quantum registers~$\sfS'_1$,~$\sfS'_2$,~$\sfT'_1$,~and~$\sfT'_2$
  possessed by the source nodes~$s_1$~and~$s_2$
  and the target nodes~$t_1$~and~$t_2$, respectively,
  are used at the stage of sharing the cat states.
  Overall, a total of seven qubits of communication are necessary
  to transfer the state from the source to the target registers.
  \label{fig:butterfly}
}
\end{center}
\end{figure}
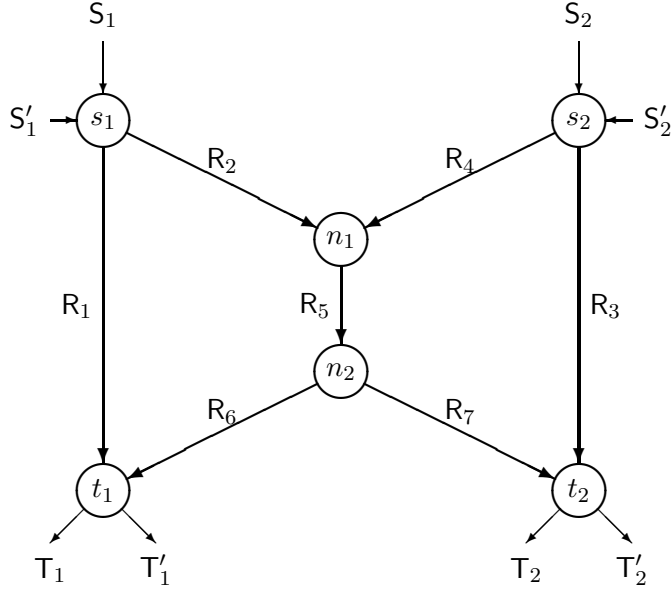

This section illustrates the techniques developed in the previous section
with the example of the quantum network shown in Figure~\ref{fig:butterfly}.
The topology of this network is the same as the classical butterfly network
(see Figure~\ref{fig:classicalbutterfly})
with the main difference that
each edge represents a quantum channel of capacity one.
Recall that in our model classical communication is free.
The task is to send a quantum state from the source ($s_1$~and~$s_2$)
to the target ($t_1$~and~$t_2$).
In this example, there are two internal nodes~$n_1$~and~$n_2$.
The difficulty is that the order of the target qubits are part of the input,
i.\,e., we have to realize
either the association corresponding to the pairs~${(s_1, t_1)}$~and~${(s_2, t_2)}$
or the association corresponding to the pairs~${(s_1, t_2)}$~and~${(s_2, t_1)}$.
The former corresponds to the identity permutation and the latter to the swap,
if we think of the qubits in some fixed order.

This task can be achieved perfectly, i.\,e., with fidelity one,
using the protocol given in Theorem~\ref{th:quantumNC}.
We give the explicit details for this example of the butterfly network.
More precisely, we describe how the protocol simulates
the classical linear coding scheme for multi-casting
presented in Figure~\ref{fig:classicalbutterfly}.
The protocol applies the fan-out operations at nodes~$s_1$,~$s_2$,~and~$n_2$,
while performs appropriate quantum coding operations at nodes~$n_1$,~$t_1$,~and~$t_2$.
Hereafter, all the registers are assumed
to be single-qubit registers each initialized to $\ket{0}$. 

First, the source node~$s_1$ (resp.~$s_2$) prepares a register~$\sfS'_1$ (resp.~$\sfS'_2$)
and applies an Hadamard operator to it.
The quantum state after this step is described as
\[
\frac{1}{2} (\ket{0}+\ket{1})_{\sfS'_1} \otimes (\ket{0}+\ket{1})_{\sfS'_2}
=
\frac{1}{2} (\ket{0}_{\sfS'_1} \ket{0}_{\sfS'_2} + \ket{0}\ket{1} + \ket{1}\ket{0} + \ket{1}\ket{1}).
\]
Then $s_1$ (resp.~$s_2$) further introduces two registers~$\sfR_1$~and~$\sfR_2$ (resp.~$\sfR_3$~and~$\sfR_4$),
and applies the operators $\CNOT^{(\sfS'_1, \sfR_1)}$ and $\CNOT^{(\sfS'_1, \sfR_2)}$
(resp.~$\CNOT^{(\sfS'_2, \sfR_3)}$ and $\CNOT^{(\sfS'_2, \sfR_4)}$).
The resulting state is
\[
\frac{1}{2}
(
  \ket{\bmzero}_{(\sfS'_1, \sfR_1, \sfR_2)} \ket{\bmzero}_{(\sfS'_2, \sfR_3, \sfR_4)}
  +
  \ket{\bmzero}\ket{\bmone}
  +
  \ket{\bmone}\ket{\bmzero}
  +
  \ket{\bmone}\ket{\bmone}
).
\]
Hereafter, let $\bmzero$ and $\bmone$ denote strings of all-zero and all-one,
respectively, of appropriate length (three here).
The registers~$\sfR_1$~and~$\sfR_2$ are sent to $t_1$ and $n_1$, respectively,
while $\sfR_3$~and~$\sfR_4$ are sent to $t_2$ and $n_1$, respectively.

The node~$n_1$ then prepares a new register~$\sfR_5$,
and applies the operators~$\CNOT^{(\sfR_2, \sfR_5)}$~and~$\CNOT^{(\sfR_4, \sfR_5)}$.
The resulting state is
\[
\frac{1}{2}
(
  \ket{\bmzero}_{(\sfS'_1, \sfR_1, \sfR_2)} \ket{\bmzero}_{(\sfS'_2, \sfR_3, \sfR_4)} \ket{0}_{\sfR_5}
  +
  \ket{\bmzero}\ket{\bmone}\ket{1}
  +
  \ket{\bmone}\ket{\bmzero}\ket{1}
  +
  \ket{\bmone}\ket{\bmone}\ket{0}
),
\]
and the register~$\sfR_5$ is sent to $n_2$. 

The node~$n_2$ then performs a quantum fan-out,
i.\,e., prepares two registers~$\sfR_6$~and~$\sfR_7$
and applies the operators~$\CNOT^{(\sfR_5, \sfR_6)}$~and~$\CNOT^{(\sfR_5, \sfR_7)}$.
The resulting state is
\[
\frac{1}{2}
(
  \ket{\bmzero}_{(\sfS'_1, \sfR_1, \sfR_2)} \ket{\bmzero}_{(\sfS'_2, \sfR_3, \sfR_4)} \ket{\bmzero}_{(\sfR_5, \sfR_6, \sfR_7)}
  +
  \ket{\bmzero}\ket{\bmone}\ket{\bmone}
  +
  \ket{\bmone}\ket{\bmzero}\ket{\bmone}
  +
  \ket{\bmone}\ket{\bmone}\ket{\bmzero}
),
\]
and the registers~$\sfR_6$~and~$\sfR_7$ are sent to $t_1$ and $t_2$, respectively.

At this point, the node~$s_1$ has the register $\sfS'_1$;
$s_2$ has $\sfS'_2$;
$n_1$ has $\sfR_2$~and~$\sfR_4$;
$n_2$ has $\sfR_5$;
$t_1$ has $\sfR_1$~and~$\sfR_6$;
$t_2$ has $\sfR_3$~and~$\sfR_7$.
Finally, $t_1$~(resp.~$t_2$) prepares two registers~$\sfT_1$~and~$\sfT'_1$~(resp.~$\sfT_2$~and~$\sfT'_2$),
and applies the operators~$\CNOT^{(\sfR_1, \sfT_1)}$,~$\CNOT^{(\sfR_1, \sfT'_1)}$~and~$\CNOT^{(\sfR_6, \sfT'_1)}$~(resp.~$\CNOT^{(\sfR_3, \sfT_2)}$,~$\CNOT^{(\sfR_3, \sfT'_2)}$~and~$\CNOT^{(\sfR_7, \sfT_2)}$).
The resulting state is
\[
\begin{split}
\hspace{5mm}
&
\hspace{-5mm}
\frac{1}{2}
(
  \ket{\bmzero}_{(\sfS'_1, \sfR_1, \sfR_2)} \ket{\bmzero}_{(\sfS'_2, \sfR_3, \sfR_4)} \ket{\bmzero}_{(\sfR_5, \sfR_6, \sfR_7)} \ket{0,0}_{(\sfT_1, \sfT'_1)} \ket{0,0}_{(\sfT_2, \sfT'_2)}
\\
&
  +
  \ket{\bmzero}\ket{\bmone}\ket{\bmone} \ket{0,1} \ket{0,1}
  +
  \ket{\bmone}\ket{\bmzero}\ket{\bmone} \ket{1,0} \ket{1,0}
  +
  \ket{\bmone}\ket{\bmone}\ket{\bmzero} \ket{1,1} \ket{1,1}
).
\end{split}
\]

Now every qubit in ${(\sfR_1, \sfR_2, \sfR_3, \sfR_4, \sfR_5, \sfR_6, \sfR_7)}$
is measured in the Hadamard basis.
The outcome~${\bmy_0 \in \Binary^7}$ is then communicated to the target node~$t_1$.
Using the information of $\bmy_0$,
the state can be mapped to
\[
\begin{split}
\hspace{5mm}
&
\hspace{-5mm}
\frac{1}{2}
(
  \ket{0}_{\sfS'_1} \ket{0}_{\sfS'_2} \ket{0,0}_{(\sfT_1, \sfT'_1)} \ket{0,0}_{(\sfT_2, \sfT'_2)}
  +
  \ket{0} \ket{1} \ket{0,1} \ket{0,1}
  +
  \ket{1} \ket{0} \ket{1,0} \ket{1,0}
  +
  \ket{1} \ket{1} \ket{1,1} \ket{1,1}
)
\\
&
\otimes
\bigl(
  H^{\otimes 7} \ket{\bmy_0}_{(\sfR_1, \sfR_2, \sfR_3, \sfR_4, \sfR_5, \sfR_6, \sfR_7)}
\bigr)
\end{split}
\]
by a local operation at $t_1$.

The state in ${(\sfR_1, \sfR_2, \sfR_3, \sfR_4, \sfR_5, \sfR_6, \sfR_7)}$
is then discarded.
Observe that
the state in ${(\sfS'_1, \sfT_1, \sfT_2, \sfS'_2, \sfT'_1, \sfT'_2)}$
in this order of the registers
forms two cat states
\[
\frac{1}{2}
(\ket{0,0,0} + \ket{1,1,1})_{(\sfS'_1, \sfT_1, \sfT_2)}
\otimes
(\ket{0,0,0} + \ket{1,1,1})_{(\sfS'_2, \sfT'_1, \sfT'_2)}.
\]

From these two cat states,
two EPR pairs can be created easily
either in ${(\sfS'_1, \sfT_1)}$ and ${(\sfS'_2, \sfT'_2)}$
or in ${(\sfS'_1, \sfT_2)}$ and ${(\sfS'_2, \sfT'_1)}$,
according to the two possible communication scenarios. 
For instance, an EPR pair in ${(\sfS'_1, \sfT_2)}$ shared by $s_1$ and $t_2$
can be created as follows.
The node~$t_1$ measures the qubit in $\sfT_1$
in the Hadamard basis~$\{\ket{+},\ket{-}\}$
and sends the result ${b \in \Binary}$ of the measurement to $s_1$.
The node~$s_1$ then applies the operator~$\sigma_Z^b$
to the qubit in $\sfS'_1$.
It can be checked easily that the remaining two qubits in ${(\sfS'_1, \sfT_2)}$ form an EPR pair.

Finally, using these EPR pairs
either in ${(\sfS'_1, \sfT_1)}$ and ${(\sfS'_2, \sfT'_2)}$
or in ${(\sfS'_1, \sfT_2)}$ and ${(\sfS'_2, \sfT'_1)}$,
the quantum state~$\rho_S$ in ${(\sfS_1, \sfS_2)}$
is teleported either to ${(\sfT_1, \sfT'_2)}$ or to ${(\sfT_2, \sfT'_1)}$
in this order of registers.
By appropriately applying swap operators,
the state~$\rho_S$ is recovered either in ${(\sfT_1, \sfT_2)}$ or in ${(\sfT_2, \sfT_1)}$
in this order of registers.


\section{Conclusions}

It has been proved that the problem of teleporting an unknown quantum state
through a network can be solved perfectly, i.\,e., with fidelity one,
by efficiently using the idea of network coding.
The method presented in this paper allows the state to be teleported
for all quantum networks
whenever classical linear network coding is possible for the network. 
Moreover, it only uses Clifford operations
and is based on three simple rules
that are applied at each node of the network:
fan-out operations, quantum coding operations, and measurements.


\subsection*{Acknowledgements}

The authors are grateful to Tsuyoshi~Ito for helpful discussions.


\end{document}